\documentclass[a4paper,fleqn]{cas-dc}

\usepackage[numbers]{natbib}

\usepackage{natbib}
\usepackage{enumerate}
\usepackage{booktabs}
\usepackage{todonotes}
\usepackage{soul}
\usepackage{makecell}
\usepackage{multirow}
\usepackage{adjustbox}
\usepackage{pdflscape}
\usepackage{afterpage}
\usepackage{caption}
\usepackage{paralist}
\usepackage[graphicx]{realboxes}

\usepackage{quoting,xparse}

\NewDocumentCommand{\bywhom}{m}{
  {\nobreak\hfill\penalty50\hskip1em\null\nobreak
   \hfill\mbox{\normalfont(#1)}%
   \parfillskip=0pt \finalhyphendemerits=0 \par}%
}

\NewDocumentEnvironment{pquotation}{m}
  {\begin{quoting}[
     indentfirst=false,
     leftmargin=\parindent,
     rightmargin=\parindent]\itshape}
  {\bywhom{#1}\end{quoting}
  }

\definecolor{mycolor}{RGB}{0, 0, 0}

\definecolor{mycolor2}{RGB}{0, 0, 0}

\definecolor{mypink1}{rgb}{0.858, 0.188, 0.478}

\begin{document}
\let\WriteBookmarks\relax
\def\floatpagepagefraction{1}
\def\textpagefraction{.001}
\title [mode = title]{Integrating Pair Programming as a Work Practice}                      

\shorttitle{Integrating PP as a Work Practice}
\shortauthors{Haugland Andersen et al.}

\author[1,2]{Nina Haugland Andersen}[orcid=0000-0002-6513-2667]
\cormark[1]
\ead{nina.haugland@ntnu.no}

\author[2,3]{Anastasiia Tkalich} [orcid=0000-0001-7391-4194]
\ead{anastasiia.tkalich@sintef.no}

\author[2,3]{Nils Brede Moe} [orcid=0000-0003-2669-0778]
\ead{nils.b.moe@sintef.no}

\author[3,2]{Darja \v{S}mite}[orcid=0000-0003-1744-3118]
\ead{darja.smite@bth.se}

\author[5]{Asgaut Mjølne Söderbom}[orcid=]
\ead{asgaut.mjolne@gmail.com}

\author[5]{Ola Hast}[orcid=]
\ead{ola.hast@gmail.com}

\author[2,4]{Viktoria Stray}[orcid=0000-0002-6032-2074]
\ead{Stray@ifi.no}
\address[1]{NTNU,
            Trondheim, Norway }
\address[2]{SINTEF,
            Trondheim, Norway}
\address[3]{Blekinge Institute of Technology,
            Karlskrona, Sweden}            
\address[4]{UIO, Oslo, Norway}   
\address[5]{Sparebank 1 Utvikling, Oslo, Norway}   

\cortext[cor1]{Corresponding author}

\nonumnote{This research was supported by the Research Council of Norway through the 10xTeams project (Grant 309344). }
\maketitle

\begin{abstract}[Summary]
\noindent
\textbf{Context:} Pair programming (PP) is more relevant than ever. As modern systems grow in complexity, knowledge sharing and collaboration across teams have become essential. However, despite well-documented benefits of PP, its adoption remains inconsistent across software teams. \textbf{Objective:} This study aims to understand the factors that facilitate or hinder team members' adoption as well as lasting engagement in PP. \textbf{Method:} We have conducted an exploratory single-case study in a mature agile company in Norway. We collected data through two rounds of interviews with team members in different roles and performed a thematic analysis of the interviews. \textbf{Results:} Our key finding is that multiple factors, related to the perceptions of how PP contributes to
daily work, efforts associated with engaging in
PP sessions, company and team attitudes, resources, infrastructure, and task characteristics, affect PP engagement. \textbf{Conclusion:} Long-term engagement in PP requires expected benefits with the practice being confirmed in firsthand experiences. Adapting the practice to each unique team, with insights drawn from collective learning, is also beneficial. Our findings will be beneficial for software practitioners seeking to make PP an integrated part of their team’s workflow.
\end{abstract}

\begin{keywords}
Pair Programming \sep Agile \sep Agile Practices \sep Empirical study
\end{keywords}

\section{Introduction and Related Work}

Pair programming (PP) has been a cornerstone practice of agile and is now even more important. As modern systems grow in complexity, knowledge sharing and collaboration are increasingly essential. In PP, two developers work together on a single computer, continuously collaborating on the same design, algorithm, code, or test \cite{williams2003pair}.
Having two minds actively engaged helps identify potential issues early and improves the correctness of complex solutions \cite{arisholm2007evaluating}, ultimately leading to more robust software. 
Further, teams strategically integrate pair programming to tackle complex features, onboard new developers, and facilitate knowledge transfer between team members. PP serves as a collaborative learning mechanism \cite{preston2005pair, vanhanen2007perceived}, fostering continuous knowledge transfer \cite{zieris2014knowledge}—a critical advantage in an era where technologies evolve rapidly.

Research has shown that pair programming enhances problem-solving and team performance \cite{kude2019pair}, with pairs consistently outperforming individual developers \cite{dybaa2008empirical,hannay2009effectiveness}. Additionally, PP contributes to employee well-being, as many developers report greater enjoyment when pairing compared to working alone \cite{williams2000strengthening, vanhanen2007perceived}. One possible reason for this is that PP aligns with key processes of effective agile teamwork, such as mutual performance monitoring, feedback, and backup behavior \cite{moe2010teamwork,salas2004}.

Despite these proven benefits and the increasing relevance of pair programming in modern software development, its adoption remains relatively low. In their study of pair programming at Microsoft, Begel and Nagappan \cite{begel2008pair} found that 22\% of participants have practiced pair programming, however, only 3.5\% applied it in their current project. Further, the Annual agile adoption surveys (2020) indicate that only 31\% of respondents report using PP in their daily work \cite{versionone2020}. 
This raises an important question: if pair programming offers such clear advantages, why is it not more widely practiced? 

Research about the adoption of agile practices in general indicates that willingness to adopt a practice is influenced by both individual and organizational elements \cite{chan2009acceptance}. Individual capabilities, experience, and thus also training play a crucial role. Further, perceived benefits, voluntariness (whether adoption is mandatory or optional), organizational culture, and active management support impact the motivation to integrate the practice in daily work. Additionally, perceived ease of use, compatibility with existing workflows, and the maturity of supporting technology can either facilitate or hinder adoption. 

When it comes to pair programming, it does not require many specific skills, knowledge or training, promises many benefits, can be considered easy to use and integrate in existing workflows, and it does not necessarily require specific technology. However, like many agile rituals, pair programming is a collaborative practice that thrives in organizational culture that supports communication and teamwork. Therefore, the pandemic-induced shift to remote work created significant obstacles to synchronous collaboration, leading to a natural decline in remote pairing \cite{smite2021collaboration, preston2005pair}. While advances in collaboration tools and the partial return to offices have mitigated some of these challenges, the widespread adoption of pair programming remains difficult, even in mature agile organizations where management actively supported its implementation.


The destiny of collaborative practices, including PP, in modern workplaces with increased remote working is a great concern. Research shows a decrease in general interest in collaborative work \cite{smite2021collaboration, kane2021redesigning}, loosening of social ties \cite{clear2021thinking}, and impaired interpersonal safety \cite{tkalich2022happens}. At the same time, collaboration, even if done virtually, is of vital importance to maintain agility and interactive exchanges among team members to prevent regression to ineffective document-centric knowledge management strategies \cite{de2022grounded}. Pairing and mutual support have demonstrated startling results in addressing inefficiencies of fully remote ~\cite{clear2021thinking} and hybrid working ~\cite{tkalich2023pair}. Thus, understanding whether adoption of collaborative practices has changed in today's work environments and finding ways to cultivate collaboration is of great importance. 

In this paper, we investigate the complex interplay of factors that facilitate and hinder the integration of pair programming into daily workflows, driven by the following research question: 

\textbf{RQ: What factors facilitate or impede the integration of pair programming as a work practice in software teams in mature agile organizations?}

Our results are based on a case study at SpareBank 1 Utvikling (SB1U), a mature agile organization in Norway. We identify critical elements related to perceived value, effort requirements, social environment, and contextual conditions that influence the willingness of the team members to continue pairing after an initial introduction. Our findings offer practical guidance for software practitioners seeking to make pair programming a natural component of their team's workflow. 

The paper is structured in the following way. Section 2 gives an introduction to SB1U as the research case. We explain the role of PP in the company and describe how it proceeded with popularizing the practice internally. Section 3 outlines our research approach and the conceptual framing. Section 4 describes the results of the data analysis. In section 5 we discuss the results in their relation to the earlier literature and summarize the practical implications of these results. Finally, Section 6 concludes the paper and suggests direction for future research.

\section{An Attempt to Popularize PP at SB1U – Why and How}

Our findings are based on the insights from SB1U where PP has been systematically encouraged as a preferred work mode. To clarify the context of our study, we will now introduce the company and describe its effort in popularizing PP.

\subsection{SB1U as a Mature Agile Environment}
SpareBank 1 Utvikling (SB1U), a Norwegian software company owned by a banking alliance, stands as a prime example of a mature agile organization that has for years focused on employee empowerment and creating an attractive workplace for developers. With 25 software teams ranging from 5 to 20 members each, SB1U has grown to 700 employees by 2022, demonstrating expansion in its digital product development capacity for web and mobile banking domains.

The company's commitment to agile practices dates back to 2012, with autonomous and highly collaborative teams at the core of their operational model. Each team typically comprises developers, testers, user experience designers, product owners, and team leaders, who have considerable freedom to decide which work methods to use. Teams commonly employ a Kanban variant incorporating elements of Scrum, along with coordination practices such as backlog meetings and daily stand-ups. Teams have adopted the weekly rituals of Monday Commitments (MC) and Friday Wins (FW). On Mondays, team members set clear intentions by committing to specific actions, ensuring focus and priority coordination for the week. By publicly committing to these actions, a sense of accountability is created among team members. On Fridays, teams reflect on and celebrate their achievements, reviewing progress relative to their MCs. This practice helps recognize progress, identify obstacles, and reinforce positive behavior, thereby maintaining momentum and engagement \cite{wodtke2017radical}. 

A pivotal transformation in SB1U's journey was its strategic shift from a traditional monolithic architecture to microservices. This architectural change was not merely technical; it fundamentally enabled a significant level of team autonomy by making teams end-to-end responsible for their products, minimizing handovers, and implementing DevOps practices effectively.

The company's commitment to employee empowerment and development is perhaps best illustrated by their innovative "20 percent policy". Implemented in 2018 in response to retention challenges, this policy allows developers to dedicate every Thursday to learning, testing new technologies, or improving common code. This initiative not only reduced employee turnover but also fostered a stronger sense of community through collaborative learning opportunities.

Further, emphasizing their commitment to knowledge sharing, SB1U actively supports guilds (communities of practice) where employees with similar interests can share knowledge and solve problems together. These communities particularly thrived during the designated learning days, creating natural opportunities for cross-team collaboration and innovation.

The company's dedication to empowerment extends to their performance management approach, utilizing objectives and key results (OKR) to guide work while maintaining team autonomy. Regular retrospectives and team health checks enable teams to continuously improve their practices, reflecting a mature agile mindset that values continuous learning and adaptation.

This combination of autonomous teams, modern architecture, dedicated learning time, and strong community building has transformed SB1U into one of Norway's most innovative tech companies, with their mobile banking app achieving top ratings in Apple's App Store and Google Play. 

\subsection{Popularization of Pair Programming}

Pre-pandemic, PP was used in SB1U sporadically rather than systematically and primarily by a few enthusiasts. As a result of remote working during the Covid-pandemic, many who were used to PP stopped doing it because of work being more individualized and because tools for remote PP were missing. When returning to hybrid ways of working, the use of PP was low in many teams. SB1U's attempt to popularize pair programming started with an intervention, where teams in two product areas attended seminars to get inspired to try PP or rejuvenate the interest in PP for those who had already tried it. This was a grassroots initiative from two PP enthusiasts (co-authors of this paper) who had experienced the value of PP and wished to share their experiences with their peers and inspire increased use of the practice in the company. One of the co-authors was also interviewed as a participant in this study.

The intervention in Product area A took place between January 15 and February 14, 2024, and in Product area B between October 8 and October 30, 2024. An overall idea with the intervention was that the threshold for team members to participate should be low, in the sense that the tasks required in the intervention should be aligned with the existing workflow. In general, the intervention consisted of four elements: 
\begin{enumerate} 
\item A kick-off meeting for all teams, where PP and the intervention, as well as "PP best practice", were presented through an inspirational talk; 
\item A three-week trial period dedicated to PP, where each team member was encouraged to schedule at least two PP sessions each week and participate in retrospective reflections on the success of these sessions each Friday; 
\item The possibility for support, the members of Team 4 and one appointed initiator in each team were available for the team members for support and coaching; 
\item Concluding product area retrospective, where team members discussed PP across teams on the final day of the intervention. 
\end{enumerate}

Based on the feedback received from the first intervention and to popularize pairing even more, team members from all roles, not only developers, were encouraged to participate in the second turn. To make the intervention more inclusive and broaden engagement, the term "pair working" was used alongside "pair programming". 

The interventions provided an opportunity to experiment with different pair programming approaches, such as actively testing different methods, constellations, work modes, setups, including the number of persons in the session. Many teams highlighted how time, guidance, and encouragement for experimentation helped tailor PP to their team's preferences and needs. 

Many of the participants appreciated the structured focus on PP and the recommended practices for testing. This led to insights grounded in personal experiences, and \textit{"aha moments"}, such as when realizing that rotation reduced fatigue. However, discussions about the intervention mainly centered on its practical aspects — how to better conduct PP, establish routines, and maintain focus — rather than shifting perceptions of PP's value. 

In the following, we describe our approach to collecting and analyzing the data at SB1U. 
\section{Research Approach}
To gain insight into what enables and hinders PP adoption and integration in the daily work of agile team members, we conducted an exploratory single-case study~\cite{runeson2009guidelines}. 
We collected data through interviews with team members participating in the intervention. The interviewees held different roles, including developers, team leaders, technical leaders, testers, and a designer, as detailed in Table \ref{tab:informants}. The interviews were conducted in two rounds with members from two different product areas. 

\begin{table}[th]
    \centering
    \small 
    \renewcommand{\arraystretch}{0.85} 
    \setlength{\tabcolsep}{2pt} 
    \caption{Summary of the participants}
    \label{tab:informants}
    \begin{tabular}{p{0.8cm} p{1.6cm} p{0.7cm} p{1.2cm} p{1.8cm}} 
        \toprule
        \textbf{No.} & \textbf{Role} & \textbf{Area} & \textbf{Team} & \textbf{Interview} \\
        \midrule
        I1  & Developer        & A & 1  & Individual \\
        I2  & Developer        & A & 2  & Individual \\
        I3  & Team lead        & A & 3  & Individual \\
        I4  & Developer        & A & 4  & Individual \\
        I5  & Developer        & A & 3  & Individual \\
        I6  & Team lead        & A & 1,4,5,8  & Individual \\
        I7  & Developer        & A & 5  & Individual \\
        I8  & Developer        & A & 4  & Individual \\
        I9  & Developer        & A & 6  & Individual \\
        I10 & Developer     & A & 7  & Individual \\
        I11 & Tech lead        & A & 8  & Individual \\
        I12 & Developer        & B & 9  & Group 1 \\
        I13 & Developer        & B & 9  & Group 1 \\
        I14 & Developer        & B & 9  & Group 1 \\
        I15 & Designer  & B & 9  & Individual \\
        I16 & Developer        & B & 10  & Group 2 \\
        I17 & Developer        & B & 10  & Group 2 \\
        I18 & Developer  & B & 10 & Group 2 \\
        I19 & Developer        & B & 11  & Group 2 \\
        I20 & Developer        & B & 9  & Group 2 \\
        I21 & Tester        & B & 9  & Group 3 \\
        I22 & Developer        & B & 10 & Group 3 \\
        I23 & Developer  & B & 9 & Group 3 \\
        \bottomrule
    \end{tabular}
\end{table}

First, in May 2024, we conducted 11 individual interviews with members of Product area A. These interviews were semi-structured and focused on PP as a practice, and we also asked how the intervention contributed to the adoption of PP in daily work. Our interview guide covered several key areas: (1) Background information about participants' roles and experience; (2) Experiences with pair programming as a practice, including typical sessions and perceived benefits/drawbacks; (3) The pair programming intervention, including implementation challenges and efforts the teams made to adapt PP to the workflow in their team; (4) Changes in practices after the intervention; and (5) The relationship between pair programming and pull requests. Examples of questions are: "\textit{Was there anything in your work situation that made it hard/easy to do PP?}" and "\textit{What measures did you take to make room for PP?}".

In November 2024, we conducted three focus group interviews with a total of 11 participants, as well as one individual interview. These interviews were conducted with members of Product area B. We first asked participants to describe their work and team environment, followed by open discussions around two main questions: "\textit{What do you think contributes to pair programming in your team?}" and "\textit{What do you think hinders pair programming in your team?}". We asked a few follow-up questions and facilitated discussions among the participants. 


To prepare for data analysis, we first used various automatic transcription software to transcribe all interviews, which produced reasonably good quality transcripts. We then manually reviewed and corrected the transcripts in cases where the automatic transcription was inaccurate, particularly focusing on technical terminology and unclear audio segments.

The first and second authors performed a thematic analysis of the interviews \cite{braun2006using}. All interview statements related to the research question were coded, and brief descriptions of the content of each statement were made. Then, codes with similar content were grouped into categories that constituted potential themes. We wrote memos to keep track of our reflections along the way and engaged in frequent discussions about the coding work and the coding structure. The potential themes were reviewed and refined to ensure that they accurately represented the data. The final four themes were aligned with the model proposed in the Unified Theory of Acceptance and Use of Technology (UTAUT) \cite{venkatesh2003user} that provided a good match with the structure of our findings, as we found at the late stages of the data analysis. In the following sub-section, we elaborate on how we used UTAUT in our research approach.  

\subsection{Conceptual framework}

\begin{figure*}
    \centering
    \includegraphics[width=1\textwidth]{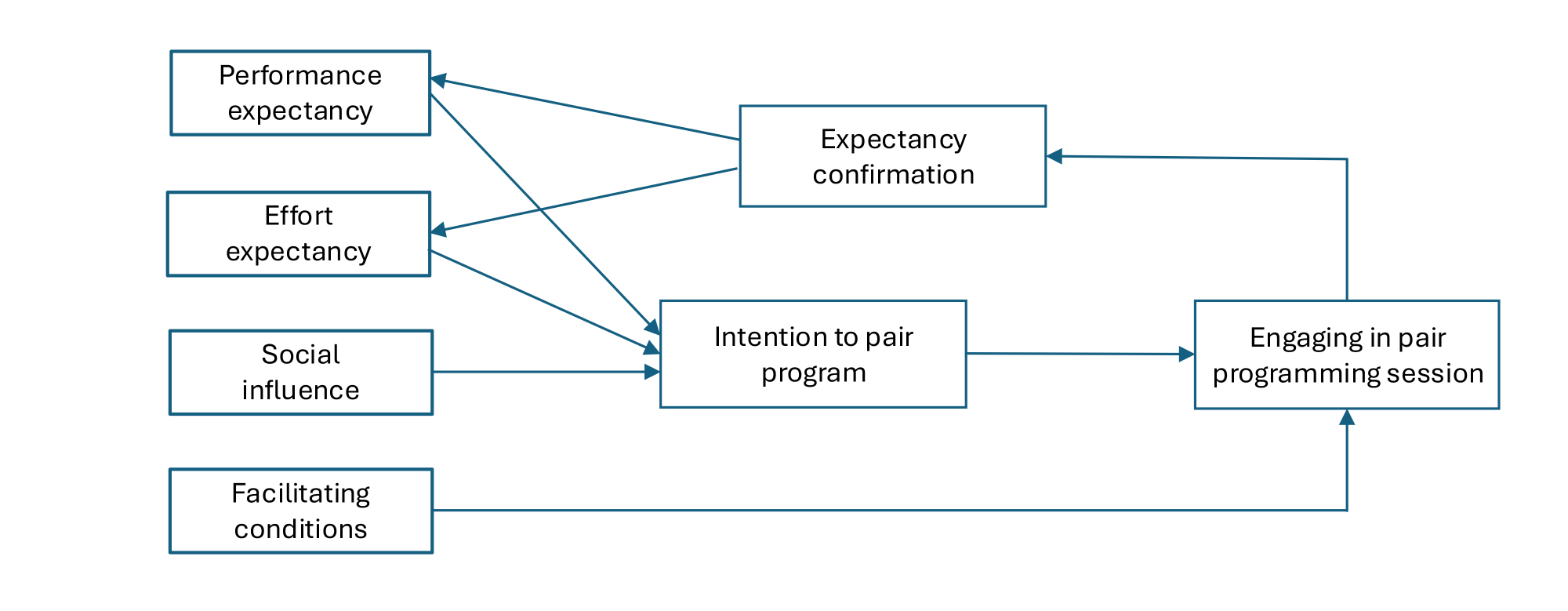}
    \caption{Initial conceptual framework} 
    \label{fig:utaut1}
\end{figure*}

In this study, we rely on a conceptual framework to investigate the integration of PP as a work practice (see Figure \ref{fig:utaut1}), which proved useful for structuring and explaining the findings during the qualitative data analysis. 

The framework was inspired by the Unified Theory of Acceptance and Use of Technology (UTAUT) \cite{venkatesh2003user} that is one of the models that has been used in the field of software engineering to explain the adoption of work practices and ways of working \cite{borstler2024acceptance}. To explain how pair programming gets integrated as a sustained practice we also drew on the concept of "expectation confirmation", which can be used together with UTAUT to explain continued use (e.g., \cite{baig2025factors}). We will now describe  the UTAUT as the main foundation of our conceptual framework and its basic concepts that will help better understand our results.  

The UTAUT model describes why someone adopts new behavior or technology or not \cite{venkatesh2003user}. Since the theory is generic, it may be applied to any practice or behavior; we will now explain how it applies specifically to pair programming. According to UTAUT, there are four main factors that determine whether someone adopts pair programming or not \cite{venkatesh2003user}: 
\begin{itemize}
    \item \textbf{Performance expectancy} - the degree to which an individual believes that PP will help him or her to attain gains in job performance. 
    \item \textbf{Effort expectancy} - the degree of ease associated with PP.
    \item \textbf{Social influence} - the degree to which an individual perceives that important others believe he or she should PP.
    \item \textbf{Facilitating conditions} - the degree to which an individual believes that objective factors in the environment exist to support PP. 
\end{itemize}

According to UTAUT, these four factors impact engagement in PP sessions. Performance expectancy, effort expectancy and social influence influence the \textit{intention} to pair program, which can then lead to actual pair programming. Furthermore, if the facilitating conditions are optimal (for example, there are suitable digital tools and support for PP throughout the organization), this alone may lead people to engage in PP sessions.   

To what extent individuals wish to \textit{continue} PP after initially trying the practice, depends on their experiences with the practice. In what ways expectations are confirmed plays an important role for continued use \cite{bhattacherjee2011information}; if the first experiences with PP confirm the positive expectations, the intention to pair program may increase, thus leading to the integration of PP in the existing workflow. However, if the first experiences are negative – negative future expectations are formed, and the intention and the likelihood to engage in PP decrease. As shown in Figure \ref{fig:utaut1}, engagement in PP produces feedback about the individual benefits and efforts associated with PP that shape the expectancy confirmation, which in terms influences future expectancies.

The results of our thematic coding according to the UTAUT model are available in Table \ref{tab:factors_pp}, which provides an account of the mentions of each of the factors influencing pair programming found in the interview data.

\section{Results}
\begin{table*}[thb]
    \centering
    \small 
    \renewcommand{\arraystretch}{0.85} 
    \setlength{\tabcolsep}{4pt} 
    \caption{Factors influencing pair programming}
    \label{tab:factors_pp}
    \begin{tabular}{p{10cm} p{3cm}} 
        \toprule
        \textbf{Factors influencing pair programming} & \textbf{N of mentions} \\
        \midrule
        \multicolumn{2}{l}{\textbf{Performance expectancy}} \\
        \hspace{5mm} Individual benefits & 50 \\
        \hspace{5mm} Team benefits & 20 \\
        \hspace{5mm} Perceived task fit & 18 \\
        \multicolumn{2}{l}{\textbf{Effort expectancy}} \\
        \hspace{5mm} Effort associated with knowledge gaps & 14 \\
        \hspace{5mm} Effort associated with feeling of fatigue & 15 \\
        \hspace{5mm} Effort associated with social interaction & 14 \\
        \hspace{5mm} Effort associated with initiating PP & 11 \\
        \multicolumn{2}{l}{\textbf{Social influence}} \\
        \hspace{5mm} Management's attitude towards the practice & 9 \\
        \hspace{5mm} Collective focus on PP in the team & 14 \\
        \multicolumn{2}{l}{\textbf{Facilitating conditions}} \\
        \hspace{5mm} Routines and measures for arranging PP sessions & 22 \\
        \hspace{5mm} Time resources & 30 \\
        \hspace{5mm} Task responsibility & 19 \\
        \hspace{5mm} Office infrastructure & 12 \\
        \hspace{5mm} Work mode & 10 \\
        \bottomrule
    \end{tabular}
\end{table*}

After the SB1U attempt to popularize pair programming through an intervention involving 11 teams in two product areas, we investigated the outcome and lasting effect of the intervention. In essence, we learned that the PP culture varied across teams. Participants from many SB1U teams described PP as natural for them, indicating that it has become integrated into daily work. One participant (I4) reinforced this by stating \textit{"[PP] is a practice that is just there"}, and another (I22) emphasized that the team is quite good at PP and that they all engage in the practice. However, this was not the case for all teams. Some participants claimed they PP less than they prefer and less than they did during the intervention. This indicates that, at least for some teams, an inspirational kick-off and a trial period alone were insufficient to make a practice stick.

Enthusiasm for PP also varies between teams. Some of the participants seem to be quite passionate about the practice and one (I6) refers to himself as a "\textit{superfan of PP}". This contrasts with the attitude expressed by a senior developer: 

\begin{pquotation}{I14}"There is not a massive support for PP specifically in our team. (...) Perhaps those of us who are seniors (...) are too used to working alone".
\end{pquotation}

This quote indicates that PP is not a habit in the team, which is necessary for PP to become a natural, integrated part of daily work practice. One participant (I2) indeed notes that habits are hard to change, and even though his team expresses willingness to PP more, they refrain from doing so in practice. As he explains: 

\begin{pquotation}{I2} "No one is stopping us (...) but you are dragged into your own tasks, and it just turns out this way. One forgets about it". 
\end{pquotation}

Our findings suggest that multiple factors affect whether individual team members engage in PP sessions. In the following subsections, we describe emotional, technological, procedural, and cultural factors that influenced engagement in PP at SB1U. The presentation of the factors follows four components of the conceptual framework: performance expectancy, effort expectancy, social influence, and facilitating conditions. (see Figure \ref{fig:utaut1}).

\subsection{Performance expectancy}
Performance expectancy - the degree to which an
individual believes that PP will help him or her to
attain gains in job performance. The participants in our study described how PP can support their individual and job performance, and detailed which tasks are perceived as suitable and which not suitable for PP. Participants in our study also highlighted that PP strengthened their well-being, which presumably has indirectly influenced their performance. 

\textbf{Individual benefits.}
If people perceived PP sessions as beneficial for them personally they seemed also more motivated to PP. The participants highlighted numerous benefits of PP, including enhanced work quality, stronger relationships with other team members, and the fact that it is pleasant and fun to work together. Several also believed that PP contributes to faster progress and increases the software development speed, particularly when compared to working alone and waiting for others to approve their pull requests. Additionally, PP sessions were described as extra support for newcomers giving an opportunity to progress without being \textit{"stuck in the dark"} (I14). On the other hand, when people did not see how they can benefit from PP, the intention to PP dropped. Few participants described that being the most proficient partner in a pair made the PP-sessions less valuable because performing this task alone would be more efficient than having to “educate” the less proficient partner.

\textbf{Team benefits.}
Whether the team could benefit from PP also mattered. Most participants agreed that PP was beneficial for knowledge-sharing in the team and one (I8) mentions that PP contributes to deploying frequently, which seems to be their motivation for PP. 

PP contribution to collective team knowledge was evident in a team that lost five developers in a short period of time, causing a permanent loss of valuable knowledge. The remaining team members realized that the redundancy of \textit{"knowing the hard parts"} is crucial (I22) and that practices such as PP help prevent such knowledge loss.


\textbf{Perceived task fit.} An important factor influencing the performance expectancy and thus engagement in PP was task suitability for PP. For example, participants who believed that "\textit{everything can be pair programmed (...) and should be pair programmed}" (I4) were among the most dedicated pair programmers. On the contrary, some perceived that certain tasks were more effectively solved individually.

The majority of participants agreed that PP provides more gains in performance when used to work on complex tasks (such as problem-solving, creating novel solutions, touching undocumented legacy code, or working on tasks involving a third party), whereas pairing on trivial tasks (such as code fidgeting) was seen as unnecessary by some. 



\begin{figure*}
    \centering
    \includegraphics[width=1\textwidth]{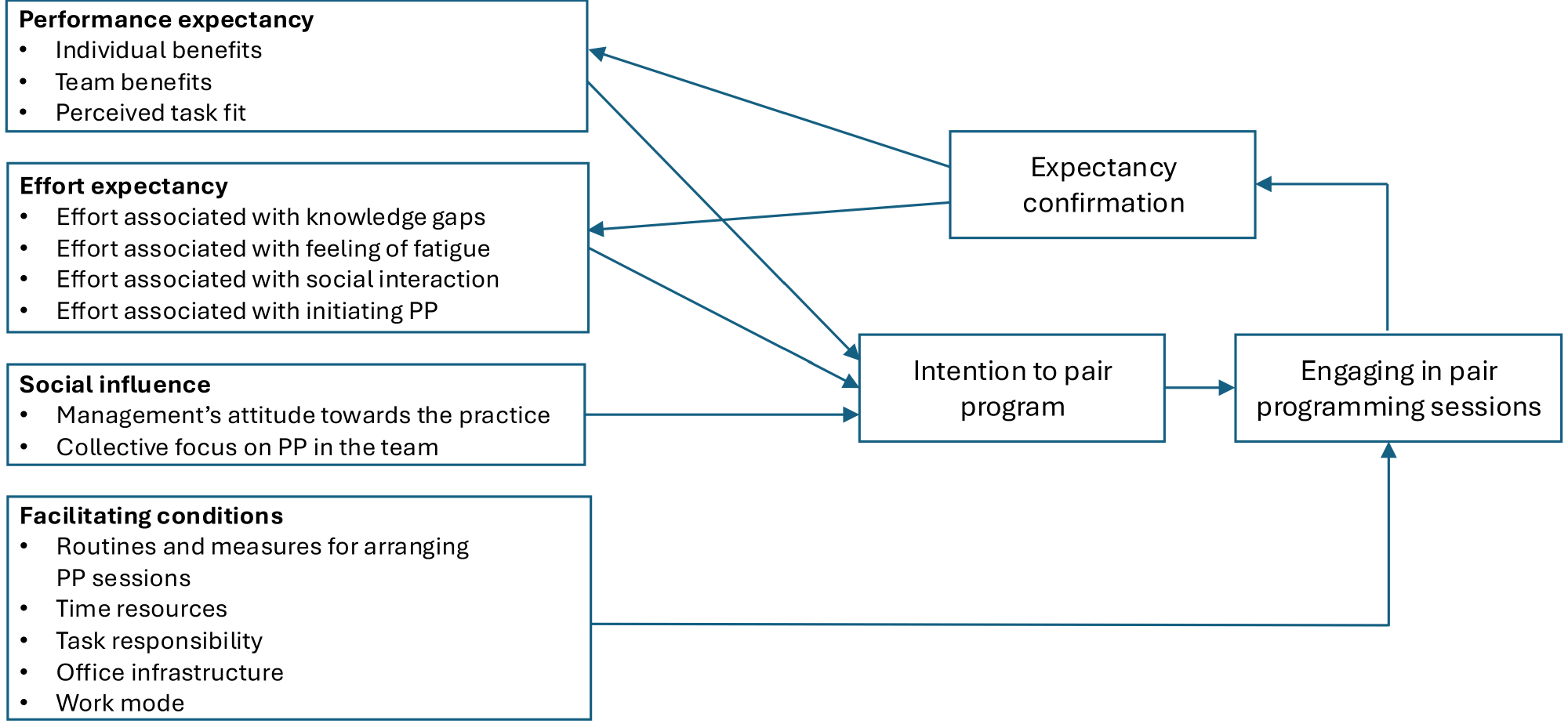}
    \caption{Summary of results}
    \label{fig:results}
\end{figure*}

\subsection{Effort expectancy}
Effort expectancy is the degree of ease associated with
PP. According to our informants, PP was not at all perceived as an easy practice. Our informants described several aspects that  
could explain why some people were more inclined to pair while others perceived PP sessions as a significant investment of their cognitive, emotional, and time resources.

\textbf{Effort associated with knowledge gaps.}
Gaps in participants' knowledge and expertise influenced the perceived cost of establishing PP sessions. We learned that participants who lacked knowledge to complete their tasks may experience cognitive strain as they must process new information while simultaneously engaging in collaborative problem-solving. Inexperienced individuals might even miss out on what is going on in the PP sessions. One participant (I23) explained that in pairs with one junior and one senior, the domain knowledge is so uneven that the junior might fail to follow what the senior does.


Therefore, those who pair with novices, inexperienced partners or partners having a different role, may need to frequently articulate their domain-specific knowledge during sessions, interrupting the flow (I13). Such sessions may result in cognitive overload, fatigue, and reluctance to pair. This is why some participants preferred to pair with someone having the same role (e.g., two front-end engineers) or with full-stack engineers. Also, one explained that PP works well when those pairing can contribute equally to problem-solving, instead of the level of knowledge is so unequal that one ends up teaching the other :

\begin{pquotation}{I7} "I think it's a sweet spot when it comes to knowledge about the apps. If one has quite equal amount of knowledge about the application, then you'll have really good discussions on how to solve things.
\end{pquotation}

Additionally, team members may feel insecure about exposing competence gaps and struggles, making them reluctant to initiate PP (I16). A key factor influencing this reluctance is psychological safety within the team, shaped by the openness to making mistakes and seeking help without fear or judgment. One participant (I15) shared that feeling psychologically safe within the team helped her to engage in PP sessions without hesitation.

\textbf{Effort associated with the feeling of 
fatigue.} The way PP sessions are held can impact how tiring they are for the pair, thus building some negative expectations and influencing their willingness to engage in PP in the future. For example, fewer breaks were perceived as more mentally exhausting compared to sessions with frequent breaks. One participant (I22) noted that the feeling of urgency to deliver often led to skipping breaks, making PP sessions more cognitively demanding. Also, being in sync when it comes to taking breaks seems to influence the cognitive load: 

\begin{pquotation}{I22} "When we are on the same page [in terms of taking breaks] one does not get so tired of pair programming".
\end{pquotation}


Interestingly, insufficient cognitive activation or boredom, which is the opposite of mental exhaustion and fatigue, was also linked to disengagement in PP-sessions. This was evident in an example brought up by a participant whose team practiced coding in triplets. The extended waiting time between turns led to a loss of focus, which was perceived negatively and reduced the willingness to engage in such sessions.

Finally, one participant mentioned that the greatest hindrance to PP was "having so fragmented tasks that they caused significant context switching" (I13), thus being cognitively demanding for the team members.



\textbf{Effort associated with social interaction.} PP sessions require not only sustained focus but also continuous social engagement, which many participants described as mentally demanding. This was especially (but not exclusively) evident among more reserved participants, who described the social aspect as draining:
\begin{pquotation}{I21}"I am about 60-70\% introvert, so the social aspect is wearing me out. Which is hard to do something about."
\end{pquotation}

On the other hand, good communication skills were found to be beneficial for PP. For example, participant I15 highlighted that clarifying expectations for mutual engagement  at the beginning of a session helps to remove tensions. Thus, it is fair to assume that the better the social skills of the partners in the PP session, the lower the effort associated with social interaction.  



\textbf{Effort associated with initiating PP.} One important precondition for habituation of PP in teams was the ease or difficulty of inviting colleagues to PP. In teams where PP was culturally ingrained, inviting others to participate required less social and emotional effort. One participant (I15) highlighted that general acceptance of PP is the most significant enabler and added that the intervention to popularize PP had lowered the threshold for initiating PP sessions. This opinion is echoed by another participant (I7), who was saying that now \textit{"leaving one's own keyboard feels more natural"}. On the contrary, in teams where PP is not widely embraced, initiating sessions demands greater social effort and can be emotionally taxing. Developers may fear rejection, skepticism, or lack of enthusiasm from their peers, increasing the psychological cost of initiating pairing. One participant (I8), who promoted PP at SB1U, explained that while management resistance has decreased, \textit{"now, the resistance comes more from developers"}. 

A few participants mentioned that there are individual team members who prefer to work alone. One participant (I5) explained that one team member had expressed a wish to work alone, therefore the two remaining team members were the ones who were left to pair. The participant added that it would have been better if all three were involved, but "\textit{it is not fun to program with a person who does not want to PP with you}" (I5). Thus, indicating that unwillingness to pair can decrease the likelihood that one is invited to PP again.

\subsection{Social influence}
Social influence is about the degree to which an individual perceives that important others believe he or she should PP. The participants highlighted how attitudes towards PP from the management and the entire team influenced their individual engagement in PP. 

\textbf{Management’s attitude towards the practice.} Perhaps not unexpectedly, the participants explain that management has a positive attitude towards PP, like expressed in a statement by I8: 
\begin{pquotation}{I8} "You experience really more and more that now nearly all leaders and team leaders want us to pair program, because they have realized how important it is". 
\end{pquotation}
The general acceptance of PP has cultivated a positive atmosphere that acts as an additional motivation to engage in PP and pairing in general.

\textbf{Collective focus on PP in the team.} One of the participants (I9) explained that the team is collectively responsible for making PP sessions a part of their workflow and that this is a precondition for PP to happen. Another (I2) agrees that PP depends on the engagement of the team members and adds that PP fades out when it does not gain focus and due attention from the members of the team. While several participants emphasized collective focus on PP as key to sustaining the practice, and the importance of persuading each other and reminding the team members about PP, others (I3) reveal that it is enough with at least a few team members actively engaging in PP to make it happen at least for a subset of the team. 


\subsection{Facilitating conditions} 
Facilitating conditions refer to the degree to which an individual percepives that objective factors in the environment exist to support PP. From the interviews, we identified several groups of organizational conditions that 
influenced PP engagement in the teams. 

\textbf{Routines and measures for arranging PP sessions.} A point brought up by a few, was that established routines and measures for arranging PP sessions, like actively deciding who will work together up front, can facilitate PP, and that the lack of such can lead to PP being done less frequently. One participant (I22) described that their team has a routine for establishing PP pairs on Mondays which they visualize on a digital board. 

We came across different routines that worked for different teams and even individuals, suggesting that there is no one-fits-all universally recommended routine for PP. For example, one participant (I23) explained that it was necessary to proactively set aside time for working with others, while another (I22) expressed no need to do so because her plans did not change so frequently and her calendar was not too busy.

\textbf{Time resources. }Limited time resources can pose difficulties in arranging PP sessions, which might lead to team members opting out. Participants who did not find the time to engage in PP mentioned competing priorities and scheduling challenges. One participant (I2) described how a planned PP session with a colleague from another team was cancelled due to more urgent tasks in the face of a deadline. Some mentioned that experienced team members, who are sought for pairing, were often unavailable due to having many booked activities. Others complained about unexpected tasks that interrupted their plans to pair, such as having to deal with a compliancy issue that national authorities had notified about. 

Additionally, participants claimed that too many tasks running in parallel resulted in the need to work solo to cover them all. However, one participant (I19) did not perceive working on many apps as a hindrance to PP, only that PP needed to happen spontaneously.

\textbf{Task responsibility. }How responsibility for tasks is distributed influences PP in a team. A few participants brought up  that having joint ownership for a task facilitates PP, as both in the pair \textit{"have the responsibility for completing [the task] together"} (I7). On the contrary, not having equal stakes in task completion creates the feeling of competition for the time resources and prioritizing one's own tasks, like expressed in the following quote: 

\begin{pquotation}{I13} "I spend a lot of time familiarizing myself with what [the others] are doing, which I could have spent on problems I feel are my responsibility".  
\end{pquotation}

One of the participants acknowledged that working together with colleagues entailed benefits for the team, like sharing knowledge about apps, but still individual responsibilities took priority, especially when working under time pressure.




\textbf{Office infrastructure.} Availability and suitability of infrastructure for PP was mentioned as a factor influencing participants' readiness to pair program. Participants brought up several issues that introduced barriers for initiating PP or immersing in the PP sessions. These were related to software, hardware, and office infrastructure. Some participants complained about curved screens being unsuitable for PP, or software tools (such as CodeWithMe) that distracted attention during the session. Noisy open landscape was also mentioned to hamper PP. 

Conversely, practical aspects can facilitate PP, such as the possibility to use one's own setup and familiar hardware. Some participants mentioned that sitting near each other in a nook or in a half-circle enables PP, for example, because one can easily roll over to each other and initiate PP (I2).

\textbf{Work location.} Another important practical consideration related to the willingness and perceived ease of participation in pair programming is the work location. During the pandemic, SBU1 employees worked remotely and experimented with tools that support pair programming. Since then, the company allows employees to work remotely a couple of days per week. However, some explain that co-location works best for PP and considerably increases the chances for initiating a PP session. There was mixed views on PP digitally. Some spoke positively about pairing when both were working from home as one avoids noise from the office, and one (I15) mentioned being uncomfortable being in the office and pairing with someone at home, as you lose perspective on how loud you speak in the office landscape. 
\section{\textcolor{mycolor}{Discussion}}

At the outset, we posed the question: If PP offers such clear advantages, why is it not more widely practiced? This study provides insights into factors helping or hindering PP from the perspective of individual members of software teams in the industry. The factors that emerged from this study are related to the perceptions of how PP contributes to daily work, the expected individual efforts associated with engaging in the practice, as well as the social environment, resources, infrastructure, and task characteristics. 

While our results echo prior research that maps factors influencing agile transitions \cite{jovanovic2020agile} and the adoption of PP \cite{Dhoodhanath}, this study also highlights the dynamic nature of the process of integrating PP in daily workflows. Specifically, we show how initial experiences, positive or negative, form certain expectations, which in turn influence the intention to pair program. These intentions materialize in actual engagement in PP sessions. This interplay of the different factors and individual intentions and engagement is illustrated in Figure \ref{fig:results}. In the following, we highlight key takeaways from our results that we believe can prove valuable in the work of integrating PP in agile teams.


\textbf{Perceiving PP as valuable is key.} Repeated positive experiences build strong positive expectancy and intentions to engage in behaviors \cite{venkatesh2003user}. Therefore, it's not surprising that we found positive perceptions of PP as a strong predictor of the practice's integration into one’s daily workflow. Our findings also confirm that PP can be beneficial in increasing both the quality and speed of work \cite{dybaa2008empirical, vanhanen2007perceived, hannay2009effectiveness}. Furthermore, we also found that PP contributes to individuals’ well-being due to collaboration and interaction with peers. This supports prior research stating that people enjoy the collaborative aspects of PP \cite{williams2000strengthening, nosek1998case, vanhanen2007perceived}. 

However, we also learned that the positive PP outcomes and enjoyment of the social aspects of PP cannot be taken for granted as these will depend on a number of factors, as we describe next. 




\textbf{PP is not effortless.} We found that, although PP may sound as a rather simple practice, with two team members being located at the same computer and alternating between roles at a regular interval, it requires substantial effort from the participants.  Our findings are in line with prior research that found pairing to be socially challenging due to incompatible pairing or personality differences \cite{Dhoodhanath} and cognitively challenging due to the intensity of the process requiring sustained energy and focus \cite{chong2007social}. In our study, we translated the challenges of under-par pairs with knowledge and experience gaps into effort expectancy that explains the psychological and emotional hindrances to pairing. 

Having negative experiences with efforts required to engage in PP can adversely impact intentions and engagement in the practice. This is especially true when experiences repeatedly confirm negative expectations, as the ways expectations are confirmed play an important role for continued engagement in a practice \cite{bhattacherjee2011information}.




\textbf{A culture welcoming PP is vital.} Aligned with prior research highlighting the important role of management support in fostering agile adoption \cite{chan2009acceptance}, we too found that management's as well as team's positive attitudes towards PP provide useful social influence. This highlights the importance of organizational culture that welcomes PP, which has the potential to reduce the perceived effort associated with the adoption and integration of this practice. 


\textbf{Tailoring solutions.} Like others who documented the mismatch between how PP is done "by the book" and how professional programmers actually pair \cite{chong2007social}, we too learned that there is no one-size-fits-all solution. How teams and pairs will practice PP is likely to be affected by a range of facilitating conditions, starting from the very availability of particular hardware, software, and physical infrastructure \cite{Dhoodhanath, chong2007social}. In our study, we found that the likelihood of succeeding with PP implementation increases if you help team members experiment and decide what is the best way to PP for their context. 



However, not all identified barriers could be fixed just by providing different hardware, software, or office accessories. In our study, we learned that PP must fit into team members' workflow as it requires simultaneous availability of several team members. Competing responsibilities among team members, in our case, made establishing regular PP sessions considerably more challenging. In a recent study, Kumar et al. \cite{kumar2025time} notes that time allocation challenge is widespread; their survey revealed that while developers would prefer to dedicate most of their time to coding, they typically spend only half of their desired time on this activity. Similarly, a recent study found that agile practitioners (non-managers) spent an average of 2.4 hours in scheduled meetings each day and had just 2.9 hours of consecutive meeting-free time \cite{stray2024behavioral}. The scarce time resources for coding translate into even more scarce resources for PP and call for changes in workload distribution and prioritization.

\subsection{Practical implications}

From this study, we offer some advice for what practitioners can do to help members in software teams to integrate PP in their daily work. To begin with, we recommend \textbf{developing and maintaining a positive culture towards PP from the management and among team members}. The company in this study had a well-established agile environment and a collaborative culture, which served as an additional motivator for the practice. However, as an organizational culture is developed and maintained through interaction \cite{sjolie_in_press} we acknowledge that culture and practice mutually influence each other; engagement in PP will contribute to shaping the collaborative culture, as well as a positive culture for PP will facilitate engagement in the practice.

It is equally important to ensure that team members gain experiences with PP. In line with literature on continued use \cite{baig2025factors,bhattacherjee2011information}, we believe that when team members gain positive experiences with PP, that assure the usefulness of PP in their daily work and strengthen their positive expectations to the practice, it will help them maintain their intention to PP. Therefore, it is pivotal to \textbf{let team members get first-hand experiences with the practice and supervise these first efforts to achieve success}. In our study, a team that lost several members in a short time recognized the crucial role of knowledge sharing, and how PP could be beneficial to achieve this. Thus, merely being told about the value of PP is insufficient.  

To popularize PP, practitioners should consider \textbf{having a trial period in which team members are actively encouraged to experiment with PP}; stimulating team members to try PP and finding the ways that suit them specifically. A trial period, for example similar to the intervention in this study, can provide the opportunities for supervising initial experiences and coaching team members who have limited or no experience with the practice. To avoid the feeling of doing "PP for PP's sake" we recommend to focus on engaging in PP to solve real tasks together. Finally, we think it is beneficial if team members are offered support and coaching during the trial period.

To keep the initial enthusiasm lasting when the trial period is over, we advise to ensure \textbf{continued focus on PP as an integral part of the organizational and team culture}, because a short trial period may be insufficient for fully integrating PP into the teams' daily work. In our study, we learned that, at least for some teams, the frequency of PP declined and, unless it becomes a habit, one could more easily forget about the practice. Therefore, continuous management and peer support is essential to motivate people to persist with PP beyond the initial trial period.

Habituation and integration of PP requires adjustments and development of individual and team routines. Purposeful team reflections contribute to improving team outcomes \cite{kneisel2020team}, which brings us to the next advice, specifically \textbf{organize team workshops with the focus on how PP can be adapted to the unique team's characteristics, context and members}. Differences between teams, such as the extent to which a team relies on working with other teams or the types of tasks the team is responsible for, can influence which measures work best for integrating the practice. In our study, it was clear that there did not seem to be any universal routines for facilitating PP, as what works for some does not necessarily work for others. Thus, dedicating time to talk about experiences gives team members the opportunity to become aware of how the practice works for them and take measures to adapt the practice. These discussions can be organized around \textit{effort expectancy} (how to minimize the effort needed to engage with the practice), \textit{facilitating conditions} (changes to routines, infrastructure, and task and time management to facilitate PP), and \textit{social influence} (actions to enhance the culture for PP).

Setting realistic expectations is clearly important. In the context of PP it is too important to \textbf{acknowledge that PP is not completely effortless, while also emphasizing the ways to mitigate efforts associated with the practice}. From this study, we learned that having frequent breaks 
reduces mental fatigue, 
that a psychologically safe climate reduces insecurities and unwillingness to demonstrate one's gaps in understanding, and good in-pair communication reduces efforts associated with social interaction. Focus on continuous PP process improvement can prevent abandoning PP after first negative experiences.

As the final piece of advice, we would like to recommend team members to \textbf{collectively assess the value of the practice}. Two (or more) heads in this case are indeed better than one. Openly discussing the benefits and drawbacks of PP (for example, during retrospectives) will help team members understand its impact on their daily work and find adjustments that may help in their particular context. Collective learning is crucial for the long-term integration of the practice in a team.

\section{Conclusions, Limitations, and Future Work}

For agile practices like pair programming (PP) to become a lasting part of a company’s workflow, these must transition from initial adoption to an integrated part of daily workflow. In this paper, we explored what helps and hinders members in software team integrate PP in their daily work at SB1U. Our key finding is that multiple factors, related to the perceptions of how PP contributes to daily work, efforts associated with engaging in PP sessions, company and team attitudes, resources, infrastructure, and task characteristics, affect PP habituation. To ensure long-term engagement in the practice, team members need first-hand positive experiences with PP that build positive expectations and intentions to continuously engage in PP. Furthermore, it can be useful for team members to actively adapt the PP practice to their unique context, with insights drawn from collective learning.

This study has certain limitations that should be taken into account when interpreting the results. Firstly, the company in this study is a mature, agile organization with a culture and members that are generally positive to collaborative practices. Thus, the experiences with habituation of PP in a company employing people skeptical to PP, and a less favorable attitude towards peer collaboration, might significantly differ. We encourage future research in such environments to complement the insights from our study. Secondly, the interviews in this study took place only a short time after the interventions. To gain a more comprehensive picture of the long-term enablers and barriers for integrating PP as a work practice, we encourage designing longitudinal studies. More work is also needed to find ways to design successful interventions to support both initial adoption and continued use of a work practice. Finally, one of the participants is also a co-author of this paper, which can potentially lead to a conflict of interest. The role of the co-authors from SB1U was to validate the case presentation and give input on the practical implications of the results. None of the industry co-authors participated in the data analysis process and therefore did not influence the final results any more than other participants.





\section*{Acknowledgements}
The authors thank the informants for sharing their experiences. This work was partially supported by the Research Council of Norway through the 10xTeams project (Grant 309344), and the companies Iterate, Zedge, and SpareBank 1 Utvikling.

\textbf{Declaration of Generative AI and AI-assisted technologies in the writing process:} During the preparation of this work the authors used AI to improve readability and language. After using AI, the authors reviewed and edited the content as needed and take full responsibility for the content of the publication.

\bibliographystyle{elsarticle-harv}




\end{document}